\documentclass{article}
\usepackage{spconf,amsmath,epsfig,amsfonts}
\usepackage{psfrag}
\usepackage{amsmath}
\usepackage{amssymb}
\usepackage{amsfonts}
\usepackage{color}
\usepackage{tabularx}
\usepackage{subfigure}

\newcolumntype{C}[1]{>{\centering\arraybackslash}p{#1}}

\newcount\hour \newcount\minute
\hour=\time \divide \hour by 60
\minute=\time
\advance \minute by \count99
\newcommand{\daytime}{%
  \ifnum\hour=0 00\else\ifnum\hour<10 0\fi\number\hour\fi:%
  \ifnum\minute<10 0\fi\number\minute%

}

\definecolor{MyGrey}{rgb}{0.5,0.5,0.5}

\usepackage{subfigure}
\renewcommand{\thesubfigure}{\thefigure.\arabic{subfigure}}
\makeatletter
\renewcommand{\p@subfigure}{}
\renewcommand{\@thesubfigure}{{\bf Fig. \thesubfigure}.\ }
\makeatother

\def\PSNR{\mathrm{ PSNR}}
\def\punit{\, \mathrm}

\title{Multiple Selection Approximation for Improved Spatio-Temporal Prediction in Video Coding}
\name{\vspace{-2mm}J\"urgen~Seiler and Andr\'e~Kaup\vspace{-1mm}}
\address{Chair of Multimedia Communications and Signal Processing, \\University of Erlangen-Nuremberg, Cauerstr. 7, 91058 Erlangen, Germany\\
{\{seiler, kaup\}@LNT.de}\vspace{-1mm}}

\begin{document}
\topmargin=0mm
\ninept
\maketitle


\begin{abstract} \label{abstract}

In this contribution, a novel spatio-temporal prediction algorithm for video coding is introduced. This algorithm exploits temporal as well as spatial redundancies for effectively predicting the signal to be encoded. To achieve this, the algorithm operates in two stages. Initially, motion compensated prediction is applied on the block being encoded. Afterwards this preliminary temporal prediction is refined by forming a joint model of the initial predictor and the spatially adjacent already transmitted blocks. The novel algorithm is able to outperform earlier refinement algorithms in speed and prediction quality. Compared to pure motion compensated prediction, the mean data rate can be reduced by up to 15$\%$ and up to 1.16 dB gain in PSNR can be achieved for the considered sequences.

\end{abstract}




\section{Introduction} \label{sec:introduction}

The transmission and playback of digital video data has become more and more popular in the past years. But the widespread usage of digital video is only possible since video sequences can be compressed efficiently. Modern hybrid video codecs as e.\ g.\ H.264/AVC \cite{Richardson2003} use several techniques to achieve this compression. One important part of hybrid video codecs is prediction which aims at extrapolating the signal parts to be encoded from already transmitted signal parts. Since only the parts of the signal are used as source for prediction that are already available at the decoder, only the difference between the predicted signal and the original signal has to be transmitted. Consequently, the amount of data to transmit and therewith the coding efficiency directly depends on the prediction quality.

For predicting the signal, current video codecs perform either a temporal or a spatial extrapolation. Spatial extrapolation is obtained by smartly continuing the signal from already transmitted regions of the actually processed frame into the region to be encoded. As described in \cite{Richardson2003}, e.\ g.\ H.264/AVC uses $13$ different modes for spatial prediction. Instead of using signal parts from the actual frame, temporal extrapolation is achieved by using previously, already completely transmitted frames. According to \cite{Dufaux1995}, these frames are used to perform motion compensated prediction for the area being encoded. In doing so, in a previous frame an area is determined that fits the signal being encoded best. The displacement between the two areas is transmitted as side information. The decoder then can use this information to select the same area from the already transmitted frame and use it for prediction. Even though most modern video codecs can intelligently switch between spatial and temporal prediction, the combined utilization of spatial and temporal redundancies for forming a predictor is only seldom performed. Very few spatio-temporal prediction algorithms exist, but as two examples for doing so, the Pixelwise Adaptive Spatio-Temporal Prediction from \cite{Day2008} and the Joint Predictive Coding from \cite{Jiang2009} should be mentioned.

In \cite{Seiler2008c}, we recently proposed another spatio-temporal prediction algorithm. This algorithm works in two stages, whereas in the first stage a preliminary temporal prediction is obtained by motion compensation. Afterwards the preliminary prediction is spatially refined by Frequency Selective Approximation (FSA) for incorporating the redundancies to the adjacent already transmitted areas into the prediction. Although this algorithm is able to improve coding efficiency significantly, it has the drawback that it is computationally very expensive. To cope with this, in \cite{Seiler2009} we proposed a modification which is noticeably faster but has a slightly decreased coding efficiency compared to the original algorithm. In the scope of this contribution, we want to propose a novel refinement step, the Multiple Selection Approximation (MSA), that combines the advantages of the two earlier algorithms. In the next subsection, the idea of spatially refined motion compensation will be reviewed and the two older algorithms will be presented briefly for pointing out their strengths and weaknesses. Afterwards the novel algorithm will be introduced, before its prospects are proven with simulations.


\section{Spatially Refined Motion Compensated Prediction} \label{sec:prediction}

\begin{figure}
	\begin{center}
		\includegraphics[width=0.35\textwidth]{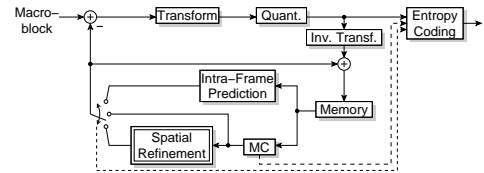}\vspace{-5mm}
	\end{center}
	\caption{\emph{Block diagram of a hybrid video encoder with spatial refinement.}}
	\label{fig:coder_diagram}
\end{figure}

\begin{figure}
	\psfrag{m}[t][t][0.8]{$m$}%
	\psfrag{n}[t][t][0.8]{$n$}%
	\psfrag{x}[t][t][0.8]{$x$}%
	\psfrag{x0}[t][t][0.7]{$x_0$}%
	\psfrag{y}[t][t][0.8]{$y$}%
	\psfrag{y0}[t][t][0.7]{$y_0$}%
	\psfrag{t}[t][t][0.8]{$t$}%
	\psfrag{t0}[t][t][0.8]{$t=\tau$}%
	\psfrag{Block}[l][l][0.8]{$\mathcal{B}$}%
	\psfrag{Rgb}[l][l][0.8]{$\mathcal{R}$}%
	\psfrag{Pgb}[l][l][0.8]{$\mathcal{P}$}%
	\centering \vspace{-7.5mm}
	\includegraphics[width=0.25\textwidth]{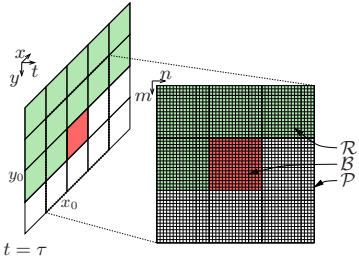}
	\caption{\emph{Relation between projection area $\mathcal{P}$ and the video sequence. $\mathcal{P}$ contains approximation area $\mathcal{R} \cup \mathcal{B}$ as union of area $\mathcal{R}$ of the reconstructed signal and block $\mathcal{B}$ to be predicted.}}
	\label{fig:prediction_area}
\end{figure}

For outlining the idea of spatially refined motion compensated prediction a common hybrid video codec operating in line scan order is regarded. In Fig.\ \ref{fig:coder_diagram}, the block diagram of such a codec is shown, including the spatial refinement step which can take place subsequent to motion compensated prediction. The block to be predicted is denoted by $\mathcal{B}$ and is located in frame $t=\tau$ at position $\left(x_0,y_0\right)$. When operating in line scan order, block $\mathcal{B}$ has four adjoining blocks that have already been transmitted and decoded. These blocks are subsumed in region $\mathcal{R}$, called reconstructed area. For the spatial refinement, now the projection area $\mathcal{P}$ of size $3\times 3$ macroblocks is regarded. As shown in Fig.\ \ref{fig:prediction_area}, area $\mathcal{P}$ is centered by the block $\mathcal{B}$ to be predicted and further contains the already transmitted blocks $\mathcal{R}$ and four padding blocks.

The spatially refined motion compensated prediction operates in two stages for forming a predictor for the signal in $\mathcal{B}$. First, motion compensated prediction is performed for this block to obtain a preliminary estimate of the signal at the expense of spending some rate for the motion vector. Afterwards, a joint model is formed for union $\mathcal{R}\cup\mathcal{B}$, the approximation area. Then, the samples corresponding to area $\mathcal{B}$ are taken from the model and are used for prediction. This joint model is used to incorporate temporal as well as spatial dependencies into the predictor and therewith form a better predictor. Since only signal parts that are also available at the decoder are used for model generation, the decoder can build the model in exactly the same way as the encoder.

The model to be generated in area $\mathcal{P}$ is denoted by $g\left[m,n\right]$ with spatial coordinates $m$ and $n$. In general, area $\mathcal{P}$ is of size \mbox{$M\times N$} samples and contains the unrefined signal $f\left[m,n\right]$ depicting the reconstructed blocks and the motion compensated block. The model
\begin{equation}
 g\left[m,n\right] = \sum_{k\in\mathfrak{K}} \hat{c}_k\varphi_k\left[m,n\right] 
  \label{eq:model}
\end{equation}
results from superimposing the mutually orthogonal two-dimen\-sional basis functions $\varphi_k\left[m,n\right]$ with corresponding weights $\hat{c}_k$. In set $\mathfrak{K}$, the indices of all basis functions used for the model generation are subsumed. So the task is to determine which basis functions to use for the model and determine the appropriate weights in such a way that the model fits the original signal to be predicted. 

The above mentioned, already proposed, algorithms for spatial refinement either utilize Frequency Selective Approximation (FSA) \cite{Seiler2008c} or Relaxed Best Approximation (RBA) \cite{Seiler2009} for the model generation. In doing so, both algorithms iteratively generate the model by selecting one (FSA) or several (RBA) basis functions per iteration to be added to the model in a certain iteration step. Thereby, the basis functions are selected in such a way that the model $g^{\left(\nu\right)}\left[m,n\right]$ in the $\nu$-th iteration step approximates $f\left[m,n\right]$, meaning that in every iteration step the weighted approximation energy
\begin{equation}
 E^{\left(\nu\right)} = \sum_{\left(m,n\right)\in \mathcal{P}} w\left[m,n\right] \left(f\left[m,n\right]-g^{\left(\nu\right)}\left[m,n\right]\right)^2
\end{equation}
is reduced. The weighting function $w\left[m,n\right]$ is used to exclude the padding blocks and to control the influence samples have on the model generation depending on their position. Samples far away from $\mathcal{B}$ in general are only weakly correlated to the signal being predicted and thus have to get a smaller weight. Therewith they have a fewer influence on the model generation compared to samples close to $\mathcal{B}$. Since the preliminary temporal extrapolation through motion compensation is an already good estimate of the original signal, this part has to get a relatively high and constant weight. According to \cite{Seiler2008c, Seiler2009} the weighting function is described by\vspace{-1mm}
\begin{equation}
 w\left[m,n\right] \hspace{-0.5mm}= \left\{\begin{array}{ll} \mu & , \forall \left(m,n\right) \in \mathcal{B} \\ \hat{\rho}^{\sqrt{\left(m-\frac{M-1}{2}\right)^2 + \left(n-\frac{N-1}{2}\right)^2}} \hspace*{-1mm}& , \forall \left(m,n\right) \in \mathcal{R} \\0 & , \mbox{ else } \end{array} \right.
 \label{eq:weighting_function}
\end{equation}
with $\mu$ controlling the weight of the motion compensated estimate and an exponentially decreasing weight for the samples in $\mathcal{R}$, controlled by decay factor $\hat{\rho}$.

\begin{figure}
	\psfrag{ph1}[c][c][0.8]{$\varphi_1$}%
	\psfrag{ph2}[c][c][0.8]{$\varphi_2$}
	\psfrag{ph3}[c][c][0.8]{$\varphi_3$}
	\psfrag{c1}[c][c][0.8]{$c_1$}%
	\psfrag{c2}[c][c][0.8]{$c_2$}%
	\psfrag{c3}[c][c][0.8]{$c_3$}%
	\psfrag{ch1}[c][c][0.8]{$\hat{c}_1$}%
	\psfrag{ch3}[c][c][0.8]{$\hat{c}_3$}%
	\psfrag{r}[l][l][0.8]{$f=c_1\varphi_1+c_2\varphi_2+c_3\varphi_3$}%
	\psfrag{a}[c][c]{a)}%
	\psfrag{b}[c][c]{b)}%
	\psfrag{c}[c][c]{c)}%
	\centering \vspace{-7.5mm}
	\includegraphics[width=0.4\textwidth]{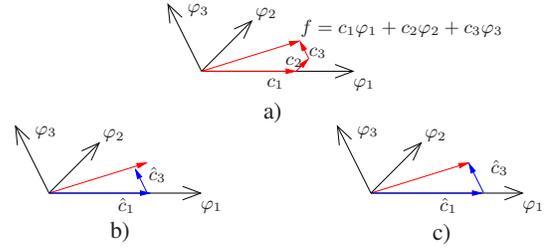}
	\caption{\emph{Approximation example:\ a) Original composition of \mbox{signal $f$}. b) Estimated coefficients after two iterations of FSA. c) Estimated coefficients after two iterations of RBA.}}
	\label{fig:od_example}
\end{figure}

For determining which basis functions to add to the model in a certain iteration step, the approximation error
\begin{equation}
 r^{\left(\nu-1\right)}\left[m,n\right] = f\left[m,n\right] -g^{\left(\nu-1\right)}\left[m,n\right]
\end{equation}
from the previous iteration step is projected onto all basis functions. In doing so, again the weighting function $w\left[m,n\right]$ is used to control the influence of the samples depending on their position. This leads for all $k$ to the projections coefficients
\begin{equation}
 p_k^{\left(\nu\right)}=\frac{\sum_{\left(m,n\right)\in\mathcal{P}}  r^{\left(\nu-1\right)}\left[m,n\right] \cdot \varphi_k\left[m,n\right] \cdot w\left[m,n\right]}{\sum_{\left(m,n\right)\in\mathcal{P}}   \varphi^2_k\left[m,n\right] \cdot w\left[m,n\right]}
 \label{eq:projection}
\end{equation}
resulting from the residual's weighted projection onto the basis function $\varphi_k\left[m,n\right]$. 

Henceforward, FSA and RBA differ. Using FSA, based on the projection coefficients, one basis function is selected to be added to the model and the portion this basis function has of the residual is added to the model and removed from the residual. But by using RBA, several basis functions are selected in every iteration step and then the input signal $f\left[m,n\right]$ is projected onto the space spanned by all the basis functions from this and the previous iterations. For this reason, the approximation of $f\left[m,n\right]$ is superior to FSA and far less iterations are needed, as shown in \cite{Seiler2009}. But it is important to notice, that the final aim is not to approximate $f\left[m,n\right]$, but to estimate the original weights of the different basis functions. For RBA, the problem arises that the estimation can fail if a basis function which is only weakly represented in the original signal is selected and added to the model. This is due to the fact, that the basis functions are not orthogonal anymore if evaluated over area $\mathcal{R}\cup\mathcal{B}$. Hence, the strongly represented basis functions that have been estimated well up to the selection of a weakly represented one, get distorted as they have to compensate the portion the weakly one has in their direction. To illustrate this problem, the two-dimensional example in \mbox{Fig.\ \ref{fig:od_example}} is regarded. There, the top subfigure shows the signal $f$, emanating from the superposition of $\varphi_1$, $\varphi_2$ and $\varphi_3$ with the original weights $c_1$, $c_2$ and $c_3$. Since the original distribution of the weights is not known, FSA and RBA aim at estimating the weights from $f$. The subfigure bottom left now shows the estimation of $\hat{c}_1$ and $\hat{c}_3$ after two iterations with FSA. And obviously, the estimated weights correspond well to the original ones, resulting in $\hat{c}_1\approx c_1$ and $\hat{c}_3\approx c_3$. In contrast to this, the subfigure bottom right shows the estimation after two iterations of RBA. There, $\hat{c}_1$ is estimated too large since $f$ is projected onto the space spanned by $\varphi_1$ and $\varphi_3$. In doing so, $\hat{c}_3$ is estimated too large, as the portion of $\varphi_2$ cannot be accounted for. And as $\varphi_1$ and $\varphi_3$ are not orthogonal, $\hat{c}_1$ has also to be enlarged to compensate the portion of $\varphi_3$ in direction of $\varphi_1$. This leads to $\hat{c}_1> c_1$ and $\hat{c}_3> c_3$.

So, although RBA achieves a better approximation the model generation is inferior to FSA, as FSA is able to estimate the weights more accurately. But at the same time, RBA needs far less iterations compared to FSA and is computationally far less expensive. Subsequently, we want to propose a novel model generation algorithm, the Multiple Selection Approximation (MSA). MSA combines the advantages of FSA and RBA, meaning that it has the same robustness against a bad basis function selection as FSA and is computationally nearly as efficient as RBA.


\section{Multiple Selection Approximation} \label{sec:musa}

The robustness of FSA is caused by the fact that in every iteration only the just selected basis function is touched and the model generated so far is not modified. Whereas the high speed of RBA originates from the circumstance that in every iteration step several basis functions are selected and the unrefined signal is projected onto the subspace spanned by all basis functions selected so far. In order to combine the advantages of both, MSA selects several basis functions per iteration and projects the residual on the subspace spanned by the basis functions selected in the actual iteration. The other basis functions, that are not involved in this iteration step, are left untouched. 

The principal behavior of MSA is akin to FSA and RBA. Again, the model from (\ref{eq:model}) is generated iteratively, whereas the initial model $g^{\left(0\right)}\left[m,n\right]$ is set to zero. To determine which basis function to select in an iteration, first of all, again a weighted projection of residual $r^{\left(\nu-1\right)}\left[m,n\right]$ is performed onto all basis functions according to (\ref{eq:projection}), yielding the projection coefficients $p_k, \forall k$. Then, the decrement 
\begin{equation}
\Delta\bar{E}_k^{\left(\nu\right)} = p_k^{\left(\nu\right)^2} \cdot\sum_{\left(m,n\right)\in\mathcal{P}} \varphi_k^2\left[m,n\right]\cdot w\left[m,n\right] 
\end{equation}
of the weighted approximation error energy that could be achieved if basis function $\varphi_k\left[m,n\right]$ alone would be removed from the residual is computed. In doing so, the basis functions that contribute strongly to the residual can be identified and a set of basis functions to use can be determined. As mentioned before, the projection on a subspace can fail if basis functions which are only weakly represented in the residual are included in the subspace. Thus, only the basis functions are selected that could lead alone to a decrement 
\begin{equation}
\Delta\bar{E}_k^{\left(\nu\right)} > \tau \max_{\tilde{k}} \Delta\bar{E}_{\widetilde{k}}^{\left(\nu\right)}
\end{equation} larger than the energy fraction threshold $\tau$ times the maximal possible decrement. In order to limit the size of the subspace, and therewith the complexity of the subsequent projection, further only the basis functions are selected that correspond to the $N_\mathrm{BF}$ largest decrements. The indices of the basis functions fullfilling both requirements then are subsumed in set $\mathfrak{G}^{\left(\nu\right)}$.

Subsequent to the basis function selection, residual $r^{\left(\nu-1\right)}\left[m,n\right]$ is projected on the space spanned by them. For this, the squared weighted distance 
\begin{equation}
 d^{\left(\nu\right)^2} \hspace{-2mm}=\hspace{-3mm} \sum_{\left(m,n\right)\in\mathcal{P}}\hspace{-1.5mm}\left(r^{\left(\nu-1\right)}\left[m,n\right] \hspace{-0.5mm}-\hspace{-2.5mm} \sum_{u\in\mathfrak{G}^{\left(\nu\right)}}\hspace{-1.5mm}\widetilde{p}_u^{\left(\nu\right)}\varphi_u\left[m,n\right]\right)^2 \hspace{-1.5mm}w\left[m,n\right]
\end{equation}
between the residual and the weighted projection onto the subspace is minimized. There, the variables $\widetilde{p}_u^{\left(\nu\right)}$ depict the new projection coefficients to determine. The minimization is carried out by setting the partial derivatives of 
\begin{equation}
\frac{\partial d^{\left(\nu\right)^2}}{\partial\widetilde{p}_u^{\left(\nu\right)}}\stackrel{!}{=}0 , \forall u \in \mathfrak{G}^{\left(\nu\right)}
\end{equation}
with respect to all $\widetilde{p}_u^{\left(\nu\right)}$ to zero. Evaluating the equations above yields the following system of $|\mathfrak{G}^{\left(\nu\right)}|$ equations which has to be solved for calculating the projection coefficients:
\[
  \sum_{\left(m,n\right)\in\mathcal{P}} \varphi_{\widetilde{u}}\left[m,n\right]r^{\left(\nu-1\right)}\left[m,n\right]w\left[m,n\right] = \hspace{2.9cm}
\]\vspace{-5mm}
\begin{equation}
  \sum_{u\in\mathfrak{G}^{\left(\nu\right)}}\widetilde{p}_u^{\left(\nu\right)} \hspace{-2mm}\sum_{\left(m,n\right)\in\mathcal{P}} \varphi_{\widetilde{u}}\left[m,n\right] \varphi_u\left[m,n\right]w\left[m,n\right] , \forall \widetilde{u} \in\mathfrak{G}^{\left(\nu\right)}
\end{equation}
In the equation above, the new index $\widetilde{u}$ is introduced to distinguish between the equation index and the index of the summation in each equation. As this system of equations is maximally of size $N_\mathrm{BF}$ and as the terms $\sum \varphi_{\widetilde{u}}\varphi_u w$ form a symmetric matrix, it can efficiently be solved by utilizing e.\ g.\ a Cholesky decomposition. 

The projection coefficients now can be used to estimate the weights $\hat{c}_u$ of the corresponding basis functions in the model. The weights are obtained from the projection coefficients by compensating the remaining orthogonality deficiency, resulting in
\begin{equation}
\hat{c}_u^{\left(\nu\right)} = \gamma \cdot \widetilde{p}_u^{\left(\nu\right)}, \forall u \in \mathfrak{G}^{\left(\nu\right)}.
\end{equation}
As outlined in \cite{Seiler2008} in detail, the problem with orthogonality deficiency is, that although the basis functions are orthogonal with respect to area $\mathcal{P}$, they are not orthogonal if evaluated in area $\mathcal{R}\cup\mathcal{B}$ in combination with the weighting function. Due to this orthogonality deficiency, the weight of a basis function can get overemphasized due to portions of other, not selected, basis functions aiming in a similar direction. To compensate this, the orthogonality deficiency factor $\gamma$ causes that only a fraction of the projection coefficient is taken and therewith overemphasizing a basis function is prevented. If the portion is estimated too small, the same basis function can get selected in a later iteration again. 

At the end of each iteration step, the model is updated by adding the selected basis functions:
\begin{equation}
  g^{\left(\nu\right)}\left[m,m\right] = g^{\left(\nu-1\right)}\left[m,m\right] + \sum_{u\in\mathfrak{G}^{\left(\nu\right)}} \hat{c}_u^{\left(\nu\right)}\varphi_u\left[m,n\right]
\end{equation}
The above described steps are repeated until a predefined number of iterations is reached. Finally, area $\mathcal{B}$ is cut out of $g\left[m,n\right]$ and is used for predicting the block to be coded.


\section{Simulations and Results}\label{sec:results} 

For evaluating the abilities of the novel algorithm, it was implemented into the H.264/AVC reference encoder JM10.2, Baseline Profile, Level 2.0. Thereby, motion compensation is carried out with quarter-pel accuracy with a maximum search range of $16$ samples and one reference frame is used. In order to compare the prediction quality, rate control is switched of and $10$ fixed QPs between $16$ and $43$ are used. In doing so, the first $100$ frames of the CIF-sequences ``Crew'', ``Discovery City'', ``Discovery Orient'', ``Foreman'', and ``Vimto'' are encoded in IPPP order at $30$ frames per second. 

Altogether, four different prediction algorithms are considered: first of all pure motion compensation without spatial refinement, further the spatial refinement is applied whereas the refinement is carried out either by FSA \cite{Seiler2008c}, RBA \cite{Seiler2009} or the novel MSA. Since for some blocks the spatial refinement can produce a worse predictor than pure motion compensation, the encoder has to test if spatial refinement should be applied or not. This simple rate-distortion optimization is performed by comparing the predictor to the original block in terms of the mean squared error. The decision if the refinement is applied  or not, further has to be transmitted to the decoder. For this, for every macroblock one additional bit is added as worst case estimate for the additional side information.

The basis functions used for model generation for the three refinement algorithms are the ones from the two-dimensional discrete Fourier transform. In \cite{Seiler2008c,Seiler2009}, this set of basis functions has already proven to be well suited for the model generation. Considering the weighting function, the preliminary temporally extrapolated block is weighted by $\mu=0.5$ and the neighboring blocks are weighted by the exponentially decreasing function with decay factor $\hat{\rho}=0.8$. The number of iterations used for the model generation is set to $200$ for FSA and to $4$ for RBA, whereas for RBA maximally 20 basis functions can get selected in an iteration step. For MSA, the number of iterations performed is set to $12$, the energy fraction threshold $\tau$ is set to $0.75$, the maximum number of basis functions per iteration to $N_\mathrm{BF}=20$ and the orthogonality deficiency compensation to $\gamma=0.5$. Fortunately, none of the mentioned parameters is very critical or sequence dependent. All the parameters can be varied widely without any worse impact on the prediction quality.

\begin{figure}
	\psfrag{s01}[t][t]{\color[rgb]{0,0,0}\setlength{\tabcolsep}{0pt}\begin{tabular}{c}$\punit{Rate} \left[\punit{kbit}/\punit{s}\right]$\end{tabular}}%
	\psfrag{s02}[b][b]{\color[rgb]{0,0,0}\setlength{\tabcolsep}{0pt}\begin{tabular}{c}$\PSNR \left[\punit{dB}\right]$\end{tabular}}%
	\psfrag{s04}[b][b]{}%
	\psfrag{s06}[][]{}%
	\psfrag{s07}[][]{}%
	\psfrag{s08}[l][l][0.64]{\color[rgb]{0,0,0}\hphantom{``Discovery Orient'':} MSA}%
	\psfrag{s21}[l][l][0.64]{\color[rgb]{0,0,0}``Discovery City'':\newlength{\ndc}\settowidth{\ndc}{``Discovery City'':}\hspace{-\ndc}\hphantom{``Discovery Orient'':} MC}%
	\psfrag{s22}[l][l][0.64]{\color[rgb]{0,0,0}\hphantom{``Discovery Orient'':} FSA}%
	\psfrag{s23}[l][l][0.64]{\color[rgb]{0,0,0}\hphantom{``Discovery Orient'':} RBA}%
	\psfrag{s24}[l][l][0.64]{\color[rgb]{0,0,0}\hphantom{``Discovery Orient'':} MSA}%
	\psfrag{s25}[l][l][0.64]{\color[rgb]{0,0,0}``Discovery Orient'': MC}%
	\psfrag{s26}[l][l][0.64]{\color[rgb]{0,0,0}\hphantom{``Discovery Orient'':} FSA}%
	\psfrag{s27}[l][l][0.64]{\color[rgb]{0,0,0}\hphantom{``Discovery Orient'':} RBA}%
	\psfrag{s28}[l][l][0.64]{\color[rgb]{0,0,0}\hphantom{``Discovery Orient'':} MSA}%
	\psfrag{x12}[t][t][0.9]{$0$}%
	\psfrag{x13}[t][t][0.9]{$500$}%
	\psfrag{x14}[t][t][0.9]{$1000$}%
	\psfrag{x15}[t][t][0.9]{$1500$}%
	\psfrag{x16}[t][t][0.9]{$2000$}%
	\psfrag{x17}[t][t][0.9]{$2500$}%
	\psfrag{x18}[t][t][0.9]{$3000$}%
	\psfrag{v12}[r][r][0.9]{$28$}%
	\psfrag{v13}[r][r][0.9]{}
	\psfrag{v14}[r][r][0.9]{$32$}%
	\psfrag{v15}[r][r][0.9]{}
	\psfrag{v16}[r][r][0.9]{$36$}%
	\psfrag{v17}[r][r][0.9]{}
	\psfrag{v18}[r][r][0.9]{$40$}%
	\psfrag{v19}[r][r][0.9]{}
	\psfrag{v20}[r][r][0.9]{$44$}%
	\psfrag{v21}[r][r][0.9]{}

	\centering \vspace{-0.6cm}
	\includegraphics[width=0.48\textwidth]{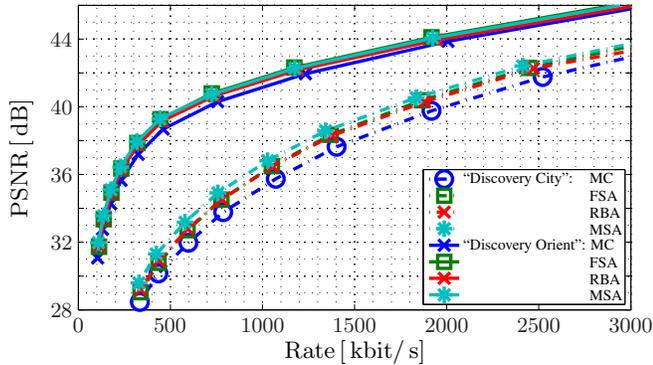}\vspace{-2mm}
	\caption{\emph{Rate-distortion curves for the first $99$ P-frames of the CIF-sequences ``Discovery City'' and ``Discovery Orient''. Comparison between unrefined motion compensation and refinement by FSA \cite{Seiler2008c}, RBA \cite{Seiler2009} and the proposed MSA.}}
	\label{fig:discovery_results}
\end{figure}

Fig. \ref{fig:discovery_results} shows the rate-distortion curves for sequences ``Discovery City'' and ``Discovery Orient'' for the case that the four different prediction algorithms are applied. Obviously, the coding efficiency can be significantly improved by applying spatial refinement subsequent to motion compensated prediction, independently of the actual refinement algorithm. In order to compare the performance of the different refinement algorithms for the regarded sequences, Table \ref{tab:results} lists the mean rate reduction and the mean $\PSNR$ gain compared to pure motion compensation respectively. The mean rate reduction and the mean $\PSNR$ gain are computed over the complete QP range according to \cite{Bjontegaard2001}. Regarding Table \ref{tab:results}, one can see that the rate can be reduced by up to $15.8\%$ and  the $\PSNR$ can be improved by up to $1.16\punit{dB}$ compared to motion compensated prediction. By further comparing the different refinement algorithms among each other, it becomes obvious that MSA is able to outperform the other two algorithms and can achieve a gain of up to $0.42\punit{dB}$ compared to FSA and up to $0.48\punit{dB}$ compared to RBA.
 
Table \ref{tab:results} further shows the mean processing time per frame for the spatial refinement. Thereby the spatial refinement is carried out in MATLAB v7.6 on one core of an Intel Core2 @ $2.4\punit{GHz}$. To accomplish this, the motion compensated block and the neighboring blocks are transferred from the JM reference software to MATLAB, where the refinement is carried out. After refinement, the processed block is retransferred to the H.264/AVC encoder. Comparing the processing time of the three algorithms, one can see that MSA is nearly as fast as RBA and $12$ times faster than FSA. The speed-up of MSA compared to FSA is slightly smaller than the ratio of the iterations, since by using MSA every iteration is a little bit more complex due to the projection onto the subspace. But overall, the reduction of the number of iterations leads to the significant reduction in processing time. Altogether, the novel refinement algorithm is able to combine the strengths of the other two algorithms: it is nearly as fast as RBA and at the same time is able to sustain or rather even improve the higher prediction quality of FSA. 

\begin{table}\vspace{-0.1cm}
\begin{center}
\scriptsize
\begin{tabular}{p{0.9cm}|C{0.8cm}|C{1.1cm}|C{1.1cm}|C{1.1cm}|C{0.8cm}}
& ``Crew'' & ``Discovery & ``Discovery & ``Foreman'' & ``Vimto''\\
& & City'' & Orient'' & &\\ \hline \hline
\multicolumn{5}{l}{Mean rate reduction}\\ \hline
MSA & $7.69\%$ & $15.84\%$ & $13.41\%$ & $2.26\%$ &  $14.98\%$ \\ \hline
FSA \cite{Seiler2008c} & $7.32\%$ & $10.38\%$ & $12.03\%$ & $3.20\%$ & $13.42\%$ \\ \hline
RBA \cite{Seiler2009} & $6.20\%$ & $9.57\%$ & $8.73\%$ & $1.42\%$ & $12.61\%$ \\ \hline
\multicolumn{5}{l}{Mean $\PSNR$ gain}\\ \hline
MSA & $0.39\punit{dB}$ & $1.16\punit{dB}$ & $0.61\punit{dB}$ & $0.09\punit{dB}$ & $0.74\punit{dB}$  \\ \hline
FSA \cite{Seiler2008c} & $0.37\punit{dB}$ & $0.74\punit{dB}$ & $0.55\punit{dB}$ & $0.13\punit{dB}$ & $0.66\punit{dB}$ \\ \hline
RBA \cite{Seiler2009} & $0.31\punit{dB}$ & $0.68\punit{dB}$ & $0.39\punit{dB}$ & $0.06\punit{dB}$ & $0.62\punit{dB}$ \\ \hline
\multicolumn{5}{l}{Mean processing time per frame}\\ \hline
MSA & $18.5\punit{s}$ & $18.2\punit{s}$ & $17.3 \punit{s}$ & $17.5\punit{s}$ & $17.5\punit{s}$  \\ \hline
FSA \cite{Seiler2008c} & $219.1\punit{s}$ & $223.3\punit{s}$ & $215.9 \punit{s}$ & $211.7\punit{s}$ & $215.9\punit{s}$  \\ \hline
RBA \cite{Seiler2009} & $13.9\punit{s}$ & $11.7\punit{s}$ & $11.3 \punit{s}$ & $12.3\punit{s}$ & $12.2\punit{s}$  \\ 
\end{tabular}
\end{center}\vspace{-5.5mm}
\caption{\emph{Mean rate reduction and mean $\PSNR$ gain compared to pure motion compensation and mean processing time per frame for spatial refinement.}}
\label{tab:results}
\end{table}


\section{Conclusion} \label{sec:conclusion} 

In the scope of this paper we presented a novel spatio-temporal prediction algorithm for video coding as improvement to two existing ones. The novel algorithm combines the advantages of the two older ones, leading to an improved prediction quality by simultaneously reducing the processing time. Compared to classical motion compensated prediction, a rate reduction of up to $15.8\%$ and a $\PSNR$ gain of up to $1.16\punit{dB}$ can be achieved for the considered sequences. The novel algorithm outperforms the better one of the two older algorithms by up $0.42\punit{dB}$ in $\PSNR$ and is nearly as fast as the faster one of the two at the same time.

Although the novel algorithm shows a significant improvement to the two older ones, our current research aims at further reducing the computational complexity.


\renewcommand{\baselinestretch}{0.80}

\end{document}